\documentstyle[11pt]{article}
\begin{document}

\begin{titlepage}
\centerline{\large\bf On the Geometry Origin of Weak CP Phase }
\vspace{1cm}

\centerline{ Yong Liu }
\vspace{0.5cm}
\centerline{Laboratory of Numeric Study for Heliospheric Physics (LHP)}
\centerline{Chinese Academy of Sciences}
\centerline{P. O. Box 8701, Beijing 100080, P.R.China}
\vspace{0.2cm}
\centerline{E-mail: soilyongliu@yahoo.com  \hspace{0.5cm}
 yongliu@ns.lhp.ac.cn}
\vspace{2cm}

\centerline{\bf Abstract}
\vspace{0.6cm}
In this work, the postulation that weak CP phase originates in a
certain geometry, is further discussed. Some intrinsic and strict
constraints on the mixing angles, weak CP phase, and the
Wolfenstein's parameters $\rho$ and $\eta$ are given by present
data and the postulation itself. Especially, we predict $0.0076
\leq |V_{td}| \leq 0.0093, \; 74.9^o \leq
\gamma \leq 75.7^o$
when the corresponding inputs are at the $90\% \ C. \ L.$. The
comparison of the predictions to the relevant experimental and
theoretical results is listed. All the predictions coincide with
the present experimental results and theoretical analysis very
well.
\\\\
PACS number(s): 11.30.Er, 12.10.Ck, 13.25.+m

\end{titlepage}

Quark mixing and CP violation
\cite{jhchr,llcha,eapas,cpvlw,cpvcj,apich,vgibs} is one of the most
interesting and important problem in particle physics. Up to now,
the origin of CP violation is not clear to us. In the standard
model, CP violation originates from a phase which presents in the
CKM matrix \cite{ncabi,mkoba}. Mathematically,
$\frac{(N-1)(N-2)}{2}$ phases are permitted to present in $N$ by
$N$ unitarity matrix in addition to $\frac{N(N-1)}{2}$ Euler
angles. However, physics is always in favour of more concise
theory.

In the previous work \cite{jlche}, we have postulated that, the
weak CP phase originates in a certain geometry. Here, we discuss
further this issue. The central purpose is to make some predictions
which can be tested by more precise data in B-factory in following
few years.

To make this paper selfcontained, we begin with describing our
postulation firstly.

\vspace{0.5cm}
\centerline{\bf 1. The Postulation}

\centerline{\it A. CKM matrix in KM parametrization and $SO(3)$
rotation}

There are many parametrization of the CKM matrix, such as the
standard one advocated by the Particle Data Group
\cite{ccaso,wykeu} and those given by Wolfenstein \cite{lwolf} etc.
However, the original parametrization chosed by Kobayashi and
Maskawa is more helpful to our understanding the problem, it is
\cite{mkoba}
\begin{equation}
V_{KM}= \left (
\begin{array}{ccc}
   c_1 & -s_1c_3& -s_1s_3 \\
   s_1c_2 & c_1c_2c_3-s_2s_3e^{i\delta}& c_1c_2s_3+s_2c_3e^{i\delta}\\
   s_1s_2 & c_1s_2c_3+c_2s_3e^{i\delta}& c_1s_2s_3-c_2c_3e^{i\delta}
\end{array}
\right )
\end{equation}
with the standard notations $s_i=\sin\theta_i$ and
$c_i=\cos\theta_i$. Note that, it is just the phase $\delta$ in
$V_{KM}$ violates CP symmetry. And all the three angles $\theta_1,
\theta_2, \theta_3$ can be taken to lie in the first quadrant by
adjusting quark field phases. In following discussions, we will fix
the three angles in first quadrant.

It is easy to find that, the above matrix can be decomposed into a
product of three Eulerian rotation matrice and one phase matrix
\cite{jfdon}.
\begin{equation}
V_{KM}= \left (
\begin{array}{ccc}
   1 & 0 & 0 \\
   0& c_2 &-s_2 \\
   0 & s_2 & c_2
\end{array}
\right )
\left (
\begin{array}{ccc}
   c_1 &-s_1 & 0 \\
   s_1& c_1 & 0 \\
   0 & 0 & 1
\end{array}
\right )
\left (
\begin{array}{ccc}
   1 &0& 0 \\
   0& 1& 0 \\
   0 & 0 & -e^{i\delta}
\end{array}
\right )
\left (
\begin{array}{ccc}
   1 &0& 0 \\
   0& c_3& s_3 \\
   0 & -s_3 & c_3
\end{array}
\right ).
\end{equation}

From the above equation, we can see that, the phase $\delta$ is
inserted into the CKM matrix some artificially. Although it is
permitted mathematically, it is not so natural physically.

Eq.(2) can easily remind us such a fact: the $SO(3)$ rotation of a
vector. Let us describe this issue more detailed. We begin with a
special example which has been written into many group theory
textbooks. Suppose that vector $\vec{V}$ locates on $X-$axis and
parallel to $Z-$axis. Now, we move it to the $Z-$axis. There are
infinite ways to do so. Here, as a special example, we consider two
of the most special ways.

1. Rotate $\vec{V}$ round $Z-$axis, after $\theta_1=\pi/2$, it is
rotated to $Y-$axis and still parallel to $Z-$axis. Then, continue
to rotate it, but this time, the rotation is round $X-$axis. After
$\theta_2=\pi/2$, $\vec{V}$ is moved to $Z-$axis, but now, it is
anti-parallel to $Y-$axis. We denote it as $\vec{V_1}$.

2. Rotate $\vec{V}$ round $Y-$axis, after $\theta_3=\pi/2$, it is
moved to $Z-$axis directly, but it is anti-parallel to $X-$axis. We
denote it as $\vec{V_2}$.

Note that, all of the movements of the vectors decribed above are
the parallel movements along the geodesics. Now, we find that,
starting out from the same vector, through two different ways of
rotation, we obtain two different vectors. The difference is only
their direction. However, if we rotate $\vec{V_1}$ round $Z-$axis
(for more general case, it is the normal direction of the point on
which $\vec{V_1}$ and $\vec{V_2}$ stand), after $\delta=\pi/2$, we
will get the same vector as $\vec{V_2}$.

From the special example, it can be seen that, the result of twice
non-coaxial rotations can not be achieved by one rotation. They are
different from each other by a "phase" $\delta$. For more general
case, $\delta$ is given by a simple relation in spherical surface
geometry
\begin{equation}
\sin\delta=\frac{(1+\cos\theta_1+\cos\theta_2+\cos\theta_3)\sqrt{
1-\cos^2\theta_1-\cos^2\theta_2-\cos^2\theta_3+
2 \cos\theta_1 \cos\theta_2 \cos\theta_3}}{
(1+\cos\theta_1)(1+\cos\theta_2)(1+\cos\theta_3)}.
\end{equation}
The geometry meaning of the above equation is shown in Fig.(1).
$\delta$ is the solid angle enclosed by the three angles $\theta_1,
\theta_2,
\theta_3$ standing
on a same point, or the area to which the solid angle corresponding
on a unit spherical surface.

It is very important to note that, it has been realized that the
magnitude of CP violation is closely related to a certain area more
than ten years ago \cite{cpvcj}.

\vspace{0.5cm}
\centerline{\it B. Phase, geometry and the weak CP phase}

Due to Berry's famous work \cite{mvber}, the phase factor has
aroused the theoretical physicists a great interests in the past
fifteen years. People have realized that, the phase is closely
related to a certain geometry or symmetry \cite{ashap,swein}. For a
non-trivial topology, the presence of the phase factor is natural.
In quantum mechanics, The well known example is the Aharonov-Bohm
effect \cite{yahar}.

To make the readers understand easily how we reached such a
postulation
- weak CP phase as a geometry phase, let us recall a simple fact in
relativity \cite{jlcmg}\cite{pkara}.

Suppose that there are two observers $A$ and $B$, $A$ observes $B$,
$A$ gets the velocity $\vec{V}$ of $B$, $B$ observes $A$, $B$ gets
the velocity $\vec{U}$ of $A$. It is evident that, $\vec{V}= -
\vec{U}$, i.e. $\vec{V}$ anti-parallel to $\vec{U}$. However, if
the third observer $C$ presents, and $A$ and $B$ observe each other
not directly but through $C$, it will not be the above case.
Suppose $A$ observes $B$ via $C$, that is, $A$ solve the velocity
of $B$ by using the velocity of $B$ relative to $C$ and the one of
$C$ relative to $A$, $A$ gets the velocity $\vec{V^\prime}$ of $B$.
Similarly, $B$ observes $A$ via $C$, $B$ gets the velocity
$\vec{U^\prime}$ of $A$. Now, although
$|\vec{V^\prime}|=|\vec{U^\prime}|$, $\vec{V^\prime}\not=-
\vec{U^\prime}$, i.e., $\vec{V^\prime}$ and $\vec{U^\prime}$ are
not parallel to each other. An angle presents between these two
veclocity vectors. This matter is illustrated in Fig.(2).

What can we learn from the above example?

First, the presence of the
angle is closely related to the presence of the third observer. This
is very similar to the case of quark mixing. If we only have two
generations of quark, we have no the weak CP phase, but, once we have
three generations of quark, the weak CP phase will present.
It is just this point stimulates us relating the weak CP phase to the
geometry phase.

Second, the more important issue we should realize here is that,
although the space in which the three observers exist is flat, the
velocity space is hyperboloidal. Or in other words, it is a
non-trivial topology. In such geometric spaces, the presence of the
phase is very natural \cite{ashap,pkara,jlcmg,gkhan,tfjor,rsimo}.

Now, if we notice that mathematically, $SO(3)\simeq S^2$ or
$SU(2)/U(1) \simeq SO(3)\simeq S^2$, and if the quarks have a
certain kind of $SO(3)$ hidden symmetry, then the phase can
present, and the spherical geometry relation Eq.(3) can be
obtained.

\vspace{0.5cm}
\centerline{\it C. The postulation in standard parametrization}

As described above, we postulate an ad hoc relation. It is, the
three mixing angles $\theta_1, \theta_2, \theta_3$ and the weak CP
phase $\delta$ satisfy Eq.(3).

If we use the standard parametrization \cite{ccaso,wykeu} instead
of KM parametrization Eq.(1), and correspondingly, we transform the
constraint Eq.(3) into the one expressed by $\delta_{13},\;\;
\theta_{12},\;\; \theta_{23}$ and $\theta_{13}$, then it will be
more convenient and clear for the following discussions.

The stardand parametrization is
\begin{equation}
V_{KM}= \left (
\begin{array}{ccc}
   c_{12} c_{13} & s_{12} c_{13}& s_{13} e^{-i \delta_{13}} \\
   -s_{12} c_{23}-c_{12} s_{23} s_{13} e^{i \delta_{13}} &
   c_{12} c_{23}-s_{12} s_{23} s_{13} e^{i \delta_{13}}    &
   s_{23} c_{13}\\
   s_{12} s_{23}-c_{12} c_{23} s_{13} e^{i \delta_{13}}  &
   -c_{12} s_{23}-s_{12} c_{23} s_{13} e^{i \delta_{13}} &
   c_{23} c_{13}
\end{array}
\right )
\end{equation}
with $c_{ij}=\cos\theta_{ij}$ and $s_{ij}=\sin\theta_{ij}$ for the
"generation" labels $i,j=1,2,3$. As the KM parametrization, the
real angles $\theta_{12}, \theta_{23}$ and $\theta_{13}$ can all be
made to lie in the first quadrant. The phase $\delta_{13}$ lies in
the range $0<\delta_{13}<2 \pi$. In following, we will also fix the
three angles $\theta_{12}, \theta_{23}$ and $\theta_{13}$ in first
quadrant.

The corresponding contraint on $\delta_{13}, \theta_{12}, \theta_{23}$
and $\theta_{13}$ - or, the expression of our postulation in this
parametrization is
\begin{equation}
\sin\delta_{13}=\frac{ (1+s_{12}+s_{23}+s_{13})
                       \sqrt{1-s_{12}^2-s_{23}^2-s_{13}^2+
                       2 s_{12} s_{23} s_{13}} }{(1+
                       s_{12}) (1+s_{23}) (1+s_{13})}
\end{equation}

From the transformation relation between the KM parametrization and
the standard one, we immediately find the following symmetry
between these two parametrization
$$
c_1\rightleftharpoons s_{13} \;\;
s_1\rightleftharpoons c_{13}\;\;
c_2\rightleftharpoons s_{23}\;\;
s_2\rightleftharpoons c_{23} \;\;
c_3\rightleftharpoons s_{12} \;\;
s_3\rightleftharpoons c_{12} \;\;
\delta \rightleftharpoons \delta_{13}
$$
$$
{\rm KM \;\;\; parametrization}\;\;\; \rightleftharpoons \;\;\;
{\rm Stardand \;\;\; parametrization}
$$
Then, Eq.(5) can be got by simple substitutions.

\vspace{0.5cm}
\centerline{\bf 2. Further investigation}

According to our postulation, only three elements in the CKM matrix
are independent. So, if we can determine experimentally three
elements of the CKM matrix, we should then determine the whole CKM
matrix. This assertion can be checked by reproducing the whole
matrix in certain error ranges with only three of the elements as
inputs.

The programme is:\\ $a$. For each group of given $V_{us},\; V_{ub}$
and $V_{cb}$, solve $s_{12},\;s_{23},\;s_{13}$ from the following
equations
\begin{equation}
|V_{us}|=s_{12}\; c_{13}\;\;\;\;\;\; |V_{ub}|=s_{13} \;\;\;\;\;\;
|V_{cb}|=s_{23}\; c_{13}.
\end{equation}
$b$. Substituting Eq.(5) into CKM matrix Eq.(4). Then, solve the
moduli of all the elements.\\ $c$. Let $V_{us},\; V_{ub}$ and
$V_{cb}$ vary in certain ranges. Repeat the steps $a$ and $b$.

When we let $V_{us},\; V_{ub}$ and $V_{cb}$ vary in the ranges
\cite{ccaso}
\begin{equation}
0.217 \leq V_{us} \leq 0.224 \;\;\;\;\;\; 0.0018 \leq V_{ub}
\leq 0.0045 \;\;\;\;\;\; 0.036 \leq V_{cb} \leq 0.042
\end{equation}

we get all the magnitudes of the elements as
\begin{equation}
\left (
\begin{array}{ccc}
0.9746 \sim 0.9762    &{\underline{0.217 \sim 0.224}}
&{\underline{ 0.0018
\sim 0.0045}}\\0.2168 \sim 0.2239
   &0.9737 \sim 0.9755  & {\underline{ 0.036
\sim 0.042}}\\ 0.0076 \sim 0.0093 &0.0352 \sim 0.0413  & 0.9991
\sim 0.9994
\end{array}
\right )
\end{equation}

Compare with those given by \cite{ccaso}
\begin{equation}
\left (
\begin{array}{ccc}
0.9745 \sim 0.9760  & 0.217 \sim 0.224  & 0.0018 \sim  0.0045\\
0.217 \sim 0.224   & 0.9737 \sim 0.9753  & 0.036 \sim 0.042\\ 0.004
\sim 0.013   & 0.035 \sim 0.042 & 0.9991 \sim 0.9993
\end{array}
\right )
\end{equation}
we find that, the predicted results are well in agreement with
those given by data book. We have reproduced the whole CKM matrix
succesfully with only three of its elements as inputs.

In the meantime, $|V_{td}|$ is not sensitive to the variations of
the inputs. It lies in a very narrow window with the central value
being about $\sim 0.0085$, even if we take a little more large
error ranges for the inputs. This may be taken as one of the
criterions to judge our postulation.

On the other hand, the relevant result extracted from the
experiment on $B_d^0-\overline{B_d^0}$ mixing gives \cite{ccaso}
\begin{equation}
|V_{tb}^* \cdot V_{td}|=0.0084 \pm 0.0018.
\end{equation}
we find that, the prediction about $|V_{td}|$ based on our
postulation, not only coincide with the experimental result very
well, but also gives a strict constraint.

\vspace{0.5cm}
\centerline{\bf 3. Predictions based on the postulation}

What can we extract from this postulation? And, how about the
correctness of the conclusions extracted from the postulation? Let
us list some other simple conclusions for more strict tests in
future.

\vspace{0.5cm}
\centerline{\it A. On the three mixing angles in CKM matrix}

To make $\theta_1, \theta_2 $ and $\theta_3
\;\;(0<\theta_i<\pi/2, \;\;i=1,\; 2,\; 3)$ enclose a solid angle,
the following relation among them should be satisfied.
\begin{equation}
\theta_i+\theta_j \geq \theta_k  \;\;\;\;\;\;
(i\not=j\not=k\not=i=1,2,3)
\end{equation}
Comparing Eq.(5) with Eq.(3), we find that, $\delta_{13}$ is the
solid angle enclosed by $(\pi/2 -\theta_{12}), (\pi/2-\theta_{23})$
and $(\pi/2-\theta_{13})$. Hence, for the standard parametrization,
the following relation should hold
\begin{equation}
(\frac{\pi}{2}-\theta_{ij})+(
\frac{\pi}{2}-\theta_{jk}) \geq (
\frac{\pi}{2}-\theta_{ki})  \;\;\;\;\;\;
(i\not=j\not=k\not=i=1,2,3. \;\;\;\theta_{ij}=\theta_{ji}).
\end{equation}
It can be checked that, Eqs.(11, 12) are easily satisfied by the
present experimental data \cite{ccaso}.

\vspace{0.5cm}
\centerline{\it B. On the weak phase $\delta_{13}$}

According to the geometry meaning of $\delta\;(\delta_{13})$, it is
the solid angle enclosed by $\theta_1, \theta_2, \theta_3\;
(\frac{\pi}{2}-\theta_{12}, \frac{\pi}{2}-\theta_{23},
\frac{\pi}{2}-\theta_{13})$.
So, if $0<\theta_i<\pi/2 \;\;( 0<\theta_{ij}<\pi/2 \;\;
i\not=j=1, 2, 3)$ and $
\theta_i+\theta_j>\theta_k \;\;((\frac{\pi}{2}-\theta_{ij})+
(\frac{\pi}{2}-\theta_{jk})\geq (\frac{\pi}{2}-\theta_{ki})  \;\;\;
i\not=j\not=k\not=i=1,2,3 \;\;\;\;\theta_{ij}=\theta_{ji})
$
are satisfied, then, $\delta\;\; (\delta_{13})$ only can lie in the
first quadrant. At most, one can take the solid angle
$\delta\;(\delta_{13})$ as $4 \pi-\delta \;(4 \pi-\delta_{13})$.
So, $2 \pi-\delta \; (2 \pi-\delta_{13})$ in the fourth quadrant
can also be permitted. Hence,

The second and third quadrants for $\delta\;\;(\delta_{13})$
are excluded thoroughly.

The recent analysis of Buras, Jamin and Lautenbacher \cite{ajbur}
indicates that, $\sin\delta_{13}$ likely lies in the first
quadrant.

\vspace{0.5cm}
\centerline{\it C. On the three angles in unitarity triangle }

The three angles $\alpha,\; \beta$ and $\gamma$ in the unitarity
triangle defined as \cite{vgibs}
\begin{equation}
\alpha \equiv arg(-\frac{V_{td} V_{tb}^*}{V_{ud} V_{ub}^*})\;\;\;\;\;\;
\beta \equiv arg(-\frac{V_{cd} V_{cb}^*}{V_{td} V_{tb}^*}) \;\;\;\;\;\;
\gamma \equiv arg(-\frac{V_{ud} V_{ub}^*}{V_{cd} V_{cb}^*})
\end{equation}

If a small change being made, it is easy to see that, the definited
angles are composed of squared and quartic invariants. For example,
$$
\alpha \equiv arg(-\frac{V_{td} V_{tb}^*}{V_{ud} V_{ub}^*})
=arg(-\frac{V_{td} V_{ud}^* V_{tb}^* V_{ub} }
{V_{ud} V_{ud}^* V_{ub} V_{ub}^*})
$$
the numerator in the definition is a quartic invariant
and the denominator is a product of two squared invariants.

Firstly, we can directly estimate that angle $\gamma$ is about
$\pi/2$. According to the geometric meaning of $\delta_{13}$, it is
the solid angle enclosed by $(\pi/2
-\theta_{12}), (\pi/2-\theta_{23})$ and $(\pi/2-\theta_{13})$. The
up-to-date experimental data \cite{ccaso} tells us that,
$s_{12}=0.217$ to 0.222, $s_{23}=0.036$ to 0.042, and
$s_{13}=0.0018$ to 0.0014. It means that, $\theta_{12},
\theta_{23}$ and $\theta_{13}$ are very small. Approximate to
the first order, we can take $\delta_{13}$ as the solid angle
enclosed by three right angles. So we get $\delta_{13}\sim \pi/2$.
On the other hand, based on the definition of $\gamma$ in Eq.(13)
and the form of standard parametrization Eq.(4), it is evident
that, $\gamma \sim \delta_{13}$. Finally, with no any detailed
calculation, we have got to know that, $\gamma \sim \pi/2$ when
approximate to the first order.

Now, let us investigate the three angles carefully. The programme
is similar to that in section {\bf 2}.\\ $a$. For each group of
given $V_{us},\; V_{ub}$ and $V_{cb}$, solve
$s_{12},\;s_{23},\;s_{13}$ from Eq.(6).\\ $b$. Substituting Eq.(5)
into CKM matrix Eq.(4). Then, solve all of the elements with the
results of $a$ being used.\\ $c$. Solve $\alpha,\; \beta$ and
$\gamma$ according to the definition Eq.(13).\\ $d$. Let $V_{us},\;
V_{ub}$ and $V_{cb}$ vary in certain ranges. Repeat the steps $a$,
$b$ and $c$.

We still let $V_{us},\; V_{ub}$ and $V_{cb}$ vary in the ranges
given by Eq.(7). The corresponding outputs are
\begin{equation}
72.1^0 \leq \alpha \leq 94.2^0 \;\;\;\;\;\; 10.7^0 \leq \beta
\leq 32.4^0 \;\;\;\;\;\; 74.9^0 \leq \gamma \leq 75.7^0.
\end{equation}

The recent analysis with more information such as a fit of
$B_d-\bar{B_d}$ mixing and $\epsilon$ being considered by Buras
gives \cite{ajbck}
\begin{equation}
35^0 \leq \alpha \leq 115^0 \;\;\;\;\;\;
11^0 \leq \beta \leq 27^0\;\;\;\;\;\;
41^0 \leq \gamma \leq 134^0
\end{equation}
or more strictly
\begin{equation}
70^0 \leq \alpha \leq 93^0 \;\;\;\;\;\; 19^0 \leq \beta \leq 22^0
\;\;\;\;\;\; 65^0 \leq \gamma \leq 90^0.
\end{equation}

It is easy to find, similar to $V_{td}$ in section {\bf 2}, we
obtain a more strict constraint on $\gamma$. We predict a very
narrow window for $\gamma$ with the central value about $\sim
75.3^0 $. Furthermore, all the predictions about $\alpha,\;
\beta$ and $\gamma$ coincide with the relevant analysis
\cite{vgibs,aalid,ajbck}.

\vspace{0.5cm}
\centerline{\it D. On the phases in the
case of more than three generations}

For the case of more than three generations, the number of the
independent phases is also $(n-1) (n-2)/2$, where $n$ is the number
of the generations. According to the geometry meaning of the phase,
the number of the independent phase is equal to the number of the
triangles which we can draw among $n$ points on a spherical surface
with the areas of the triangles are independent.

We consider the case of the four generations as a special example,
the geometry picture is shown in Fig.(3). In fact, we can draw four
spheric triangles $\triangle ABC, \triangle BCD,\triangle
CDA,\triangle DAB$ among the four vertexes $A, \; B,\; C$ and $D$.
But, due to the constraint $S_{\triangle ABC}+S_{\triangle
ADC}=S_{\triangle BAD}+S_{\triangle BCD}$, only three of
them are independent. In the meantime, the constraints
$S_{\triangle ABC}+S_{\triangle ADC}=S_{\triangle BAD}+S_{\triangle
BCD}$ and $\theta_{i}+\theta_{j}> \theta_{k} (i\not=j\not=k =1, \;
2,\; ..., 6.\; 0<\theta_{i}<\pi/2 \;{\rm and} \;
\theta_i,\;\theta_j,\;\theta_k
\; {\rm can\; enclose\; a\; solid\; angle)}$ give limits on the six
Euler angles in the CKM matrix.

Starting out from the geometry picture, we can extract some useful
informations about the mixing angles between the fourth and the
first three generations based on the mixing angles among the first
three generations. For instance, we have
$$\theta_{14}+\theta_{24}+\theta_{34}<\frac{3}{4} \pi
+ \frac{1}{2} (\theta_{12} +\theta_{23}+\theta_{13})$$ where, the
definitions of $\theta_{14},\;\theta_{24},\;\theta_{34}$ are
similar to $\theta_{12},\; \theta_{23},\;\theta_{13}$ in Eq.(4).
Substituting the present data \cite{ccaso}, we obtain
$\theta_{14}+\theta_{24}+\theta_{34}<142.7^o$.

\vspace{0.5cm}
\centerline{\it E. On the Wolfenstein's parameters $\eta$ and $\rho$ }

To make it be convenient to use the CKM matrix in the concrete
calculations, Wolfenstein parametrized it as \cite{lwolf}
\begin{equation}
V_{W}= \left (
\begin{array}{ccc}
   1-\frac{1}{2}\lambda^2 & \lambda & A\lambda^3(\rho-
   i\eta+i\eta\frac{1}{2}\lambda^2) \\
   -\lambda & 1-\frac{1}{2}\lambda^2-i\eta A^2 \lambda^4 &
   A\lambda^2(1+i\eta\lambda^2)\\
   A\lambda^3(1-\rho-i\eta) & -A\lambda^2 & 1
\end{array}
\right ).
\end{equation}

Actually, one can take different parametrization in different
cases. They are only for the convenience in discussing the
different questions, but the physics does not change when adopting
various parametrizations.

According to Buras etc., there is a very nice corresponding
relation between Wolfenstein's parameters and the ones in the
standard parametrization. It reads \cite{ajbrf}
\begin{equation}
s_{12}=\lambda, \;\;\;\;\;\; s_{23}=A\lambda^2, \;\;\;\;\;\;
s_{13}e^{-i \delta_{13}}=A\lambda^3 (\rho-i\eta).
\end{equation}
So,
\begin{equation} s_{13}=A\lambda^3 \sqrt{\rho^2+\eta^2},
\;\;\;\;\;\;
\sin\delta_{13}=\frac{\eta}{\sqrt{\rho^2+\eta^2}}
\end{equation}
and consequently,
\begin{equation}
\rho=\frac{s_{13}}{s_{12}s_{23}}\cos \delta_{13}, \;\;\;\;\;\;
\eta=\frac{s_{13}}{s_{12}s_{23}}\sin \delta_{13}.
\end{equation}

In Eq.(18), $\lambda$ and $A$ are the two better known parameters.
But, because of the uncertainty of hadronic matrix elements and
other reasons, it is difficult to extract more information about
$\rho$ and $\eta$ from experimental results. Up to now, we still
know little about them. More than ten years ago, Wolfenstein
estimated that the upper limit on $\eta$ is about $0.1$
\cite{lwolf}, but the recent analysis indicate that, $\rho$ and
$\eta$ are about \cite{jlros,zzxin,mschm}
\begin{equation}
-0.15< \rho <0.35, \;\;\;\;\;\;\;\;\;\; 0.20< \eta<0.45.
\end{equation}

Now that the four angles in CKM matrix are not independent, the
four Wolfenstein's parameters $A, \; \lambda,\; \rho$ and $\eta$
must also be not independent.

Substituting Eqs.(18-19) into Eq.(5), it is easy to achieve
\begin{equation}
\frac{\eta}{\sqrt{\rho^2+\eta^2}}=1-\frac{\lambda^2}{2}-A \lambda^3
+\lambda^4 (-\frac{1}{8}+A-\frac{A^2}{2}-A \sqrt{\rho^2+\eta^2}).
\end{equation}
This is just the geometry constraint on Wolfenstein's parameters
when approximate to the fourth order of $\lambda$.

In following, let us investigate carefully the permitted ranges of
$\rho$ and $\eta$ by present data. If we start out from Eq.(22)
directly, and take \cite{ccaso,ynira}
$$
\lambda=0.2196\pm0.0023 \;\;\;\;\;\; A=0.819\pm 0.035
$$
as inputs, then we can obtain the dependence of $\eta$ on $\rho$.
The result is shown in Fig.(4). It can be seen from the figure
that, $\eta$ and $\rho$ satisfy an approximate linear relation.

We can also begin with Eq.(20). But, we should know the three
mixing angles firstly. This can be achieved by use of three of the
CKM matrix elements such as $V_{us}$, $V_{ub}$ and $V_{cb}$. In
section {\bf 2}, we have found that, the whole matrix can be
reconstructed very well based only on three of the elements and
Eq.(5). Once the three mixing angles are determined, we can extract
the dependence of $\eta$ on $\rho$ again from Eq.(20). We take the
relevant inputs from the data book \cite{ccaso} as Eq.(7). The
numerical result is also shown in Fig.(4). We find it is just a
little part of that drawn from Eq.(22).

Now, we can read from the figure that, when all the three inputs
$V_{us},\; V_{ub}$ and $V_{cb}$ are taken at $90\% \; C. \ L.$, we
obtain the outputs
\begin{equation}
0.048<\rho<0.140, \;\;\;\;\;\;\;\;\;\; 0.18<\eta<0.54.
\end{equation}

Comparing with Eq.(21), the range for $\rho$ is more narrow while
the range for $\eta$ is relative wide. However, with more precise
measurement on the relevant CKM matrix elements in future, we can
determine them more accurately.

\vspace{0.5cm}
\centerline{\bf 4. Conclusions and discussions}

Based on the postulation that weak CP phase originates in a certain
geometry, some intuitive results are obtained. We summarize the
results as following.

1. There exists a intrinsic constraint on the three mixing angles
in CKM matrix. It is
$$
\theta_i+\theta_j \geq \theta_k  \;\;\;\;\;\;
(i\not=j\not=k\not=i=1,2,3)
$$
or
$$
(\frac{\pi}{2}-\theta_{ij})+(\frac{\pi}{2}-\theta_{jk}) \geq (
\frac{\pi}{2}-\theta_{ki})  \;\;\;\;\;\;
(i\not=j\not=k\not=i=1,2,3. \;\;\;\theta_{ij}=\theta_{ji})
$$

2. Predict undoubtedly that, if all the mixing angles are made to
lie in the first quadrant, the second and third quadrant for
$\delta\;(\delta_{13})$ are excluded thoroughly.

3. The $90 \% \; C. L.$ inputs $V_{us},\; V_{ub}$ and $V_{cb}$
gives
$$
72.1^0 \leq \alpha \leq 94.2^0 \;\;\;\;\; 10.7^0 \leq \beta
\leq 32.4^0 \;\;\;\;\; 74.9^0 \leq \gamma \leq 75.7^0
$$
for the unitarity triangle.

4. For the case of more than three generations, the numbers of the
independent phases is equal to that given by Kobayashi-Maskawa
theory. If the fourth generation exists, the present data gives a
limit on the mixing angles between the fourth and the first three
generations as $$\theta_{14}+\theta_{24}+\theta_{34}<142.7^o$$.

5. The constraint on the Wolfenstein's parametrization is worked
out, and the present data implies that
$$
0.048<\rho<0.140,
\;\;\;\;\;\;\;\;\;\; 0.18<\eta<0.54.
$$

We find that all the predictions coincide with the present data and
the relevant analysis very well.

Especially, the postulation gives strict constraints on the moduli
of CKM matrix element $|V_{td}|$ and the angle $\gamma$ in
unitarity triangle. They are predicted as
$$
0.0076 \leq |V_{td}| \leq 0.0093, \;\;\;\;\;\; 74.9^0 \leq
\gamma \leq 75.7^0.
$$
These results can be taken as the key criterions to judge our
postulation in next two years.

Our postulation can be further put to the more precise tests in
$B-$factory in following few years. If it can be verified finally,
it means that, only three elements in CKM matrix are independent,
and hence we can remove one of the free parameters in the standard
model. If then, we will feel that the physics is more simple,
natural, and beautiful. Besides, it can provide us some hints to
the hidden symmetry. But, we will naturally ask, what is the
dynamic origin?

Although our postulation is supported by the present experimental
results. It is a ad hoc supposition now, it still needs the further
verification by experiments and the basic theory on which it can
base. The further theoretical work on this problem is under way.

Because the CP phase can originate from many ways in different
theories and physical processes, we hope that our postulation at
least provide partial origin of the CP violation even if it is not
the whole origin. At least, it is a good parametrization for the
weak CP phase.

\vspace{0.5cm}

\noindent {\bf Acknowledgment}: The author would like
to thank Prof. E. A. Paschos for his encouragement and the referee
for his kind suggestions and criticism. This work is partially
supported by National Natural Science Foundation of China under the
grant number 49990453.

\begin{figure}[htb]
\mbox{}
\vskip 7in\relax\noindent\hskip -1 in\relax
\includegraphics{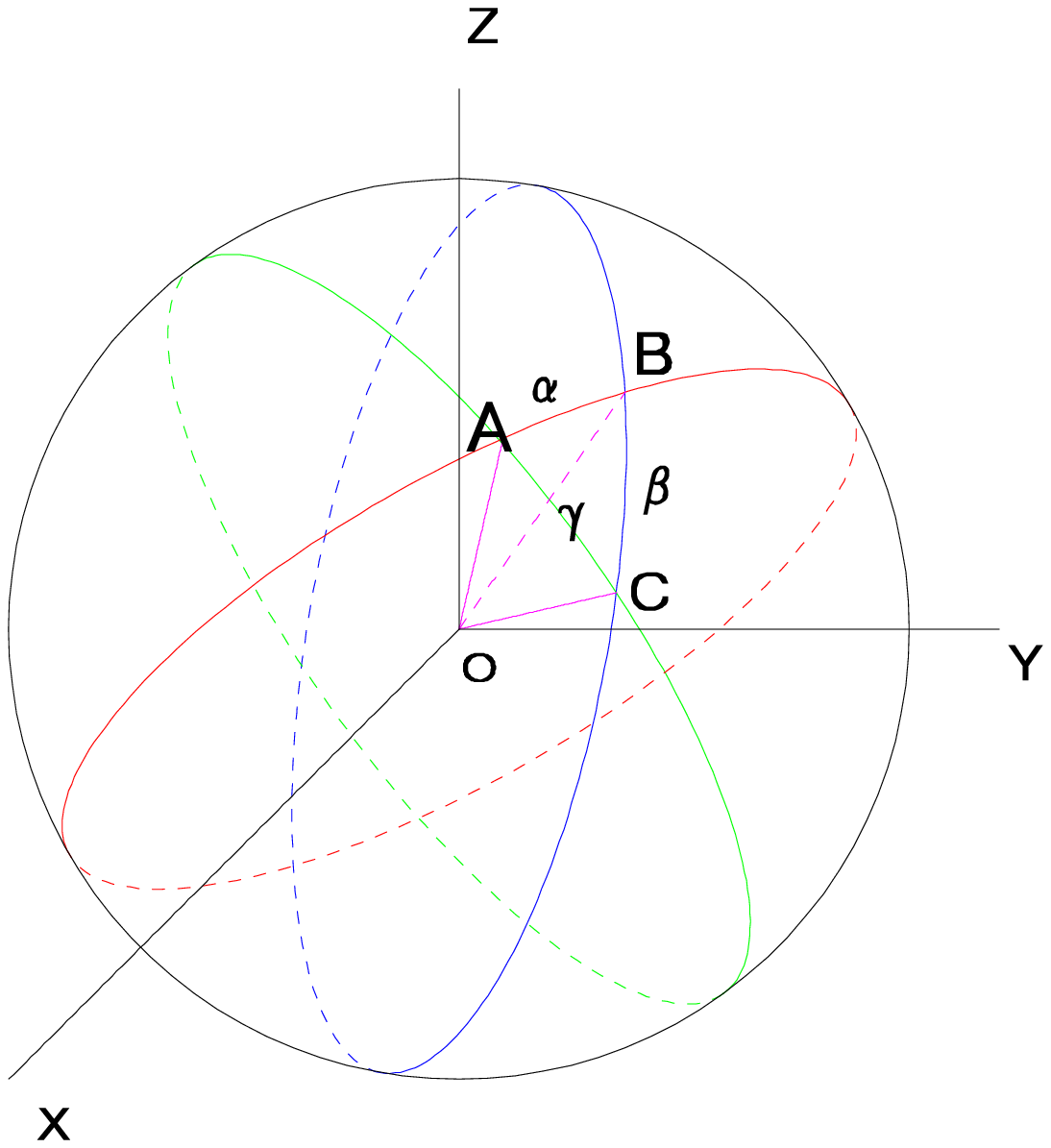}
\caption{The solid angle and the spheric triangles.
Where, $\alpha (= AOB),\; \beta(= BOC),\;
\gamma(= COA)$ represent $\theta_{1}, \; \theta_{2}, \;
\theta_{3}$. $\delta$ is the area of the spheric triangle $\triangle ABC$,
or the solid angle enclosed by the three angles $AOB,\; BOC$, and $
COA$. }
\end{figure}

\begin{figure}[htb]
\mbox{}
\vskip 7in\relax\noindent\hskip -1 in\relax
\includegraphics{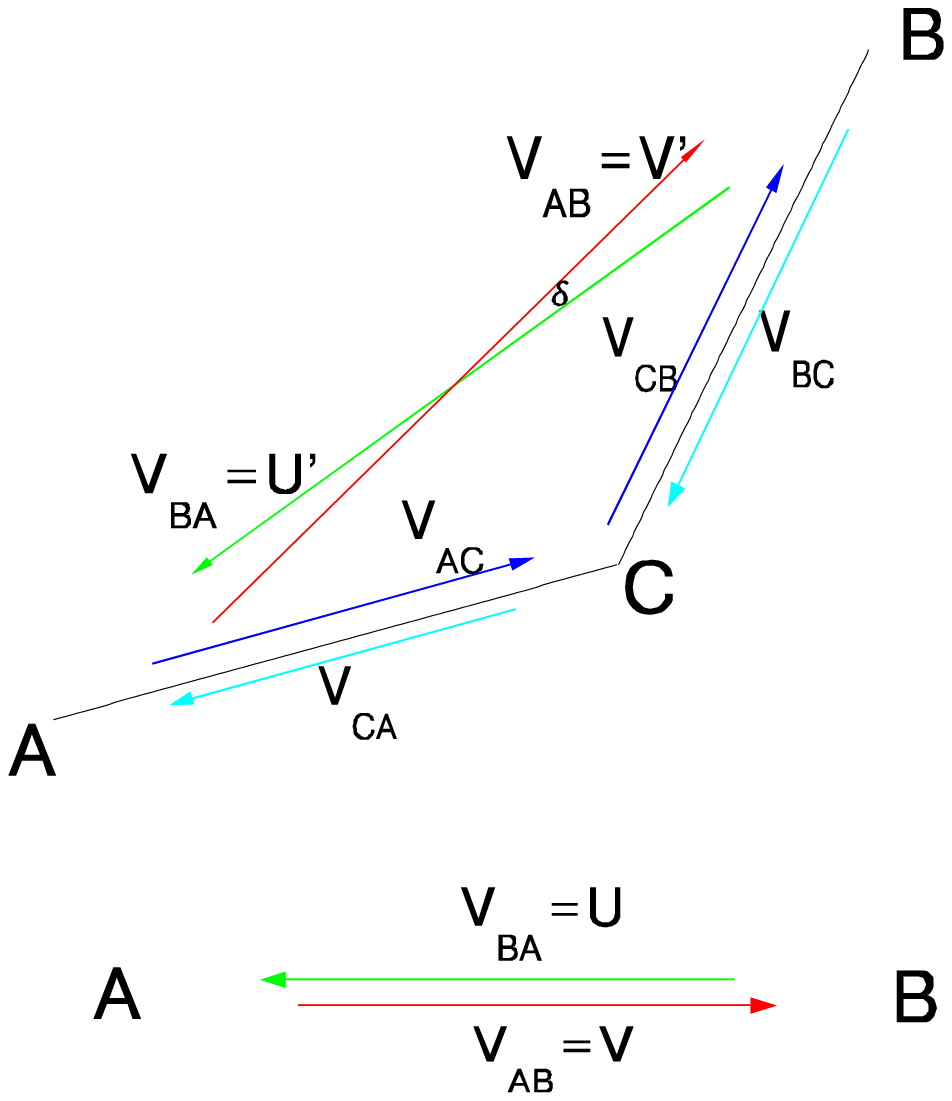}
\caption{If there are only two observers $A$ and $B$, then,
$\vec{U}=-\vec{V}$. When the third observer $C$ presents, and we
know that the velocities of $C$ relative to $A$ and the one of $B$
relative to $C$ are $\vec{V_{AC}}$ and $\vec{V_{CB}}$ respectively,
the velocities of $C$ relative to $B$ and the one of $A$ relative
to $C$ are $\vec{V_{BC}}(=-\vec{V_{CB}})$ and
$\vec{V_{CA}}(=-\vec{V_{AC}})$ respectively, then we can solve the
velocities of $B$ relative to $A$ and the one of $A$ relative to
$B$, which are denoted by $\vec{V}^\prime$ and $\vec{U}^\prime$
respectively. Now, we find that $\vec{V}^\prime \not=
-\vec{U}^\prime$, an angle $\delta$ presents between them. }
\end{figure}

\begin{figure}[htb]
\mbox{}
\vskip 7in\relax\noindent\hskip -1 in\relax
\includegraphics{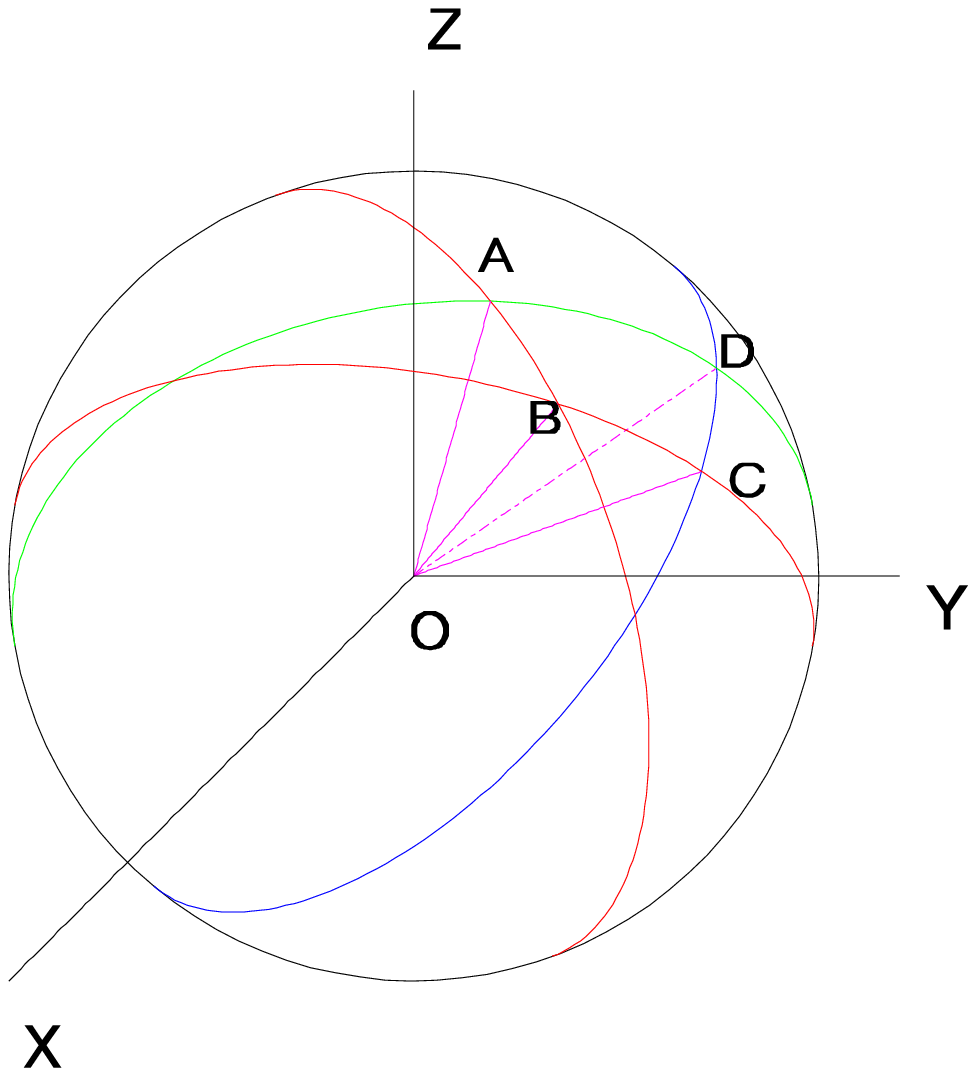}
\caption{The geometry picture for the case of the four generations.
Where, $\theta_i \;(i=1, \; 2, \; ...\; 6.)\; {\rm are\; the\;
angles }\; AOB,\; AOC,\; AOD,\; BOC,\; BOD \; {\rm and }\; COD$.
$\delta $s are the areas of the spheric triangles enclosed by any
three of the vertexes $A,\; B,\; C$ and $D$, or the solid angles
enclosed by the angles such as $AOB,\; AOC,\; BOC$ etc.. Because
$S_{ABC}+S_{CDA}=S_{BCD}+S_{DAB}$, only three of the four solid
angles (or the areas of the spheric triangles) are independent.
Here, $S_{IJK}$ represents the area of the spheric triangle
enclosed by the vertexes $I,\; J$ and $K$.}
\end{figure}

\begin{figure}[htb]
\mbox{}
\vskip 7in\relax\noindent\hskip -1 in\relax
\includegraphics{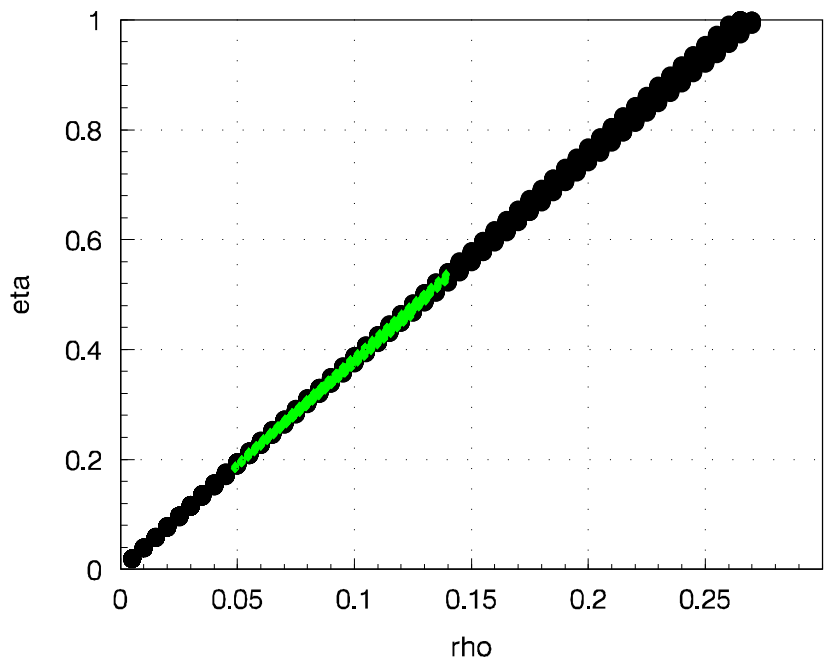}
\caption{The dependence of $\eta$ on $\rho$ based on Eq.(22) and
the permitted ranges for them by the present data. Here, the errors
of the inputs have been considered.}
\end{figure}

\end{document}